\documentclass[twocolumn,prc,showpacs,amsmath,dcolumn]{revtex4}
\usepackage{epsfig}
\begin{document}

\title{NONPERTURBATIVE RELATIVISTIC APPROACH TO PION FORM FACTOR:
PREDICTIONS FOR FUTURE JLAB EXPERIMENTS}

\author{A.F.~Krutov$^1$} \email{krutov@ssu.samara.ru},
\author{V.E.~Troitsky$^2$} \email{troitsky@theory.sinp.msu.ru}
\author{N.A.~Tsirova$^{3,1}$} \email{natalia.tsirova@clermont.in2p3.fr}

\affiliation{$^1$Samara State University, 443011 Samara, Russia}

\affiliation{$^2$D.V.~Skobeltsyn Institute of Nuclear Physics,
  Moscow State University, Moscow 119991, Russia}

\affiliation{$^3$Laboratoire de Physique Corpusculaire, 63170 Aubiere, France}
\email{natalia.tsirova@clermont.in2p3.fr}

\date{\today}
\begin{abstract}
Some predictions concerning possible results of the future JLab experiments
on the pion form factor
$F_{\pi}(Q^2)$ are made. The calculations exploit the method
proposed previously by the authors and based on the
instant--form Poincar\'e invariant approach to pion considered
as a quark--antiquark system. Long ago, this model has predicted
with surprising accuracy
the values of $F_{\pi}(Q^2)$ measured later in JLab experiment.
The results are almost independent
from the form of wave function.
The pion mean square radius $\langle r^2_{\pi} \rangle$
and the decay constant $f_{\pi}$ also agree with experimental values. The
model gives power-like asymptotic behavior of $F_{\pi}(Q^2)$ at high
momentum transfer in agreement with QCD predictions.
\end{abstract}

\maketitle

\section{INTRODUCTION}
The recent high accuracy experiments on the measurement of the pion form
factor in the range of $Q^2$ up to 2.45 GeV${}^2$ ~\cite{Blo08-1,HuB08-2}
($Q^2 =-q^2,\;q$ - momentum transfer) and future JLab experiments up to
$Q^2 \approx 6$ GeV${}^2$~\cite{Proposal06,BlH02} enhanced the interest to
theoretical description of pion at high $Q^2$ .

It is usually believed that these future experiments will provide
a meaningful test of the transition between perturbative and
non-perturbative regions which is expected at much lower $Q^2$ in
the case of pion that
of other hadrons, in
particular, of nucleon. At the present time different theoretical
approaches to the pion form factor $F_{\pi}(Q^2)$ exist.  They are
partly listed and described in Ref.~\cite{HuB08-2} (section IV) (see also
\cite{EbF06}).
In the frameworks of some of these models, a certain agreement with
existing experimental data is obtained for soft $F_{\pi}(Q^2)$. As to the
region of high momentum transfer, the theoretical results differ from one
another to a great extent. It seems us that the situation is such that one
has almost no hope to find the appearance of perturbative degrees of
freedom in the future JLab experiments on $F_{\pi}(Q^2)$. It is difficult
to imagine that in the wide band of non-perturbative theoretical curves
there would not occur any one agreeing with the experimental data. Our
opinion is that the large variety of non-perturbative predictions for
future data for $F_{\pi}(Q^2)$ makes it necessary to formulate the problem
of detecting of perturbative effects in a slightly different way than it
is usually done. Namely, we propose to accept one of the theories which
describes correctly the existing data and continue the calculations for
higher $Q^2$. If it will occur that beginning from some values one needs
to adjust the calculations by introducing the quark mass dependence on
$Q^2$ to agree the future data, then we will identify these values with
the appearance of perturbative effects. It is natural that in the present
paper we choose our own approach \cite{KrT01} as an example for the
demonstration of the proposed scenario.

The reasons for this choice are the following. The main one is that our
approach has already demonstrated its predictive power: without any
additional tuning of parameters, we
predicted in Ref.~\cite{KrT01} the values of $F_{\pi}(Q^2)$ obtained
later in experiment \cite{Vol01,Tad07,Hor06}. At the same time, the
approach gives the correct values of the mean square radius (MSR),
the decay constant $f_\pi$ and the power-like asymptotic behavior.
Certainly, other criteria of discrimination the approach may exist. For
example, one can consider as a ``correct'' one an approach which gives a
consistent treatment of the pion form factor in space-like and time-like
regions.

In the present paper, we use the approach to the pion form factor
$F_{\pi}(Q^2)$ proposed in our papers about ten years ago  \cite{KrT01}.
Our approach presents one of the versions of the constituent quark model
(CQM). The method is based on the dispersion approach to the instant form
of the Poincar\'e invariant quantum mechanics \cite{KrT02} (see also the
detailed version \cite{KrT01long} and the review  \cite{KrT09} ).

Based on this approach and on the experimental data on the measurement of
$F_{\pi} (Q^2)$
in the range of $Q^2$ up to $0.26$ (GeV)$^2$
\cite{Ame84},
we have obtained in  1998 the model function for the pion form factor for
the extended range of higher momentum transfers \cite{KrT01}. The
experimental data obtained later \cite{Vol01,Tad07,Hor06} (see also the
review of all experimental results in Ref.~\cite{HuB08-2} and references
therein) for the range of $Q^2$ larger by an order of magnitude coincide
precisely with our theoretical curve of 1998 \cite{KrT01} with no need of
any additional fitting. This means that it is possible to consider our
calculations \cite{KrT01} as an accurate prediction of the present
experimental data for the pion form factor. The model describes correctly
 the pion MSR and the decay constant $f_{\pi}$, too.      It is important
 to notice that the dependence of our results for $F_{\pi}(Q^2)$ on the
form of wave functions is very weak \cite{KrT01}. Moreover, our approach
gives the correct power-like asymptotic behavior of $F_{\pi}(Q^2)$ at $Q^2
\to \infty$. So, the model works well at  high as well as at low values of
$Q^2$.

Taking into account these advantages of our approach, it seems
natural to hope that the model will continue to give a good description of
experimental data at higher momentum transfers, in particular of the
future JLab measurements in the range $2.45 \le Q^2 \le 6 $ (GeV)$^2$
(after having withstand the test of tenfold increasing of $Q^2$ range it
may withstand another much smaller increase). If it will occur that the
experimental data does not follow our theoretical curve then we shall
adjust the theory by taking into account the quark-mass dependence on
$Q^2$. Within our approach, this depedence is a manifestation of
appearance of the perturbative degrees of
freedom.

The paper is organized in the following way. We start in Sec.~II with
a brief review of the basic theoretical formalism of our approach. The
results of calculations and the comparison with the experimental data and
other theoretical models are given in Sec.~III. In Sec.~IV the asymptotic
behavior of the form factor is considered. Finally, our conclusions are
given in Sec.~V.

\section{THE MODEL}
Our method
is a version of the instant form of the
Poincar\'e invariant constituent-quark model (PICQM),
formulated on the base of a dispersion approach
(see, e.g., \cite{KrT01,KrT02}).
As is well known the dispersion approach is based on the general
properties of space and time and, therefore, is to a certain extent ``model
independent''. That is why the calculation of electromagnetic form factors
using dispersion approach are of distinguished character as compared with
other approaches. This advantage of our method is emphasized in Ref.
\cite{DeD08} (see, however, the footnote
\footnote{Our free form factor differs from the free form factor of
Ref.~\cite{DeD08} because in \cite{DeD08} the normalizations of
one-particle
wave vectors and those of two-particle wave vectors are inconsistent in
the basis where the motion of the
two-particle center of mass is separated.
}).

The main point of our approch is the construction of the operator of
electromagnetic current which preserves Lorentz covariance and
conservation laws in the relativistic invariant impulse approximation (so
called modified impulse approximation (MIA)) \cite{KrT02}. This
approximation is constructed using dispersion-relation integrals
over composite-particle mass, that is over the Mandelstam variables
$s, s'$ \cite{KrT01long}. This variant of dispersion approach was
developed in Refs.~\cite{ShT69,KoT72,TrT72,KiT75,Tro93,AnK92,Mel02}
and was fruitfully used to investigate the structure of composite systems.

Let us recall some principal points of our approach  \cite{KrT01,KrT02}.
In
our variant of PICQM, pion electromagnetic form factor in MIA has the form
\begin{equation}
F_\pi(Q^2)=\int
\mathrm{d}\sqrt{s}\,\mathrm{d}\sqrt{s'}\,\varphi(k)\,g_0(s,Q^2,s')\,\varphi(k')\;.
\label{ffpi}
\end{equation}
Here $\varphi(k)$ is pion wave function in the sense of
PICQM, $g_0(s,Q^2,s')$ is the free two-particle form factor. It may be
obtained explicitly by the methods of relativistic kinematics and is a
relativistic invariant function.

The wave function in (\ref{ffpi}) has the following structure:
$$
\varphi(k) = \sqrt[4]{s}\,u(k)k\;,\quad s=4(k^2+M^2)\;.
$$
Here $M$ is the mass of the constituent quark.
Below for the function $u(k)$ we use some phenomenological wave functions.

The function  $g_0(s,Q^2,s')$ is written in terms of the quark
electromagnetic form factors in the form
$$
g_0(s,Q^2,s')=
  \frac{(s+s'+Q^2)Q^2}{2\sqrt{(s-4M^2)
(s'-4M^2)}}\;
$$
$$
\times \frac{\theta(s,Q^2,s')}{{[\lambda(s,-Q^2,s')]}^{3/2}}
\frac{1}{\sqrt{1+Q^2/4M^2}}
$$
$$
\times\left\{(s+s'+Q^2)[G^q_E(Q^2)+G^{\bar q} _E(Q^2)]\right.
$$
$$
\times\cos{(\omega_1+\omega_2)} + \frac{1}{M}\,\xi(s,Q^2,s')
(G^q_M(Q^2)
$$
\begin{equation}
\left.+ G^{\bar q}_M(Q^2))\sin(\omega_1+\omega_2)\right\}\;.
\label{g_0}
\end{equation}
Here $\lambda(a,b,c)=a^2+b^2+c^2-2(ab+ac+bc)$,
$$
\xi=\sqrt{ss'Q^2-M^2\lambda(s,-Q^2,s')}\;,
$$
$\omega_1$ and
$\omega_2$ are the Wigner rotation parameters:
$$
\omega_1\!=\!\arctan{\frac{\xi(s,Q^2,s')}{M[(\sqrt s\! +\! \sqrt
{s'})^2\! +\! Q^2]\! +\! \sqrt{ss'}(\sqrt s\! +\! \sqrt{s'})}},
$$
$$
\omega_2\! =\! \arctan{\frac{\alpha(s,s')\xi(s,Q^2,s')}{M(s\!
+\! s'\! +\! Q^2)\alpha( s,s')\! +\! \sqrt{ss'}( 4M^2\! +\!
Q^2)}},
$$
$\alpha(s,s') =  2M+\sqrt s+\sqrt {s'}$, $\theta(s,Q^2,s')=
\vartheta(s'-s_1) -\vartheta(s'-s_2)$, $\vartheta$ is the step
function,
$$
s_{1,2}=2M^2+\frac{1}{2M^2} (2M^2+Q^2)(s-2M^2)
$$
$$
\mp
\frac{1}{2M^2} \sqrt{Q^2(Q^2+4M^2)s(s-4M^2)}.
$$
Note that the magnetic form factor contribution to
Eq.~(\ref{g_0})  is due to the spin  rotation  effect  only
\cite{KrT99}.
Here $G^{u,\bar{d}}_{E,M}(Q^2)$ are electric and magnetic form factors of
quarks, respectively.

Let us note that we introduce electromagnetic quark form factors,
in particulary, in order to  obtain a description of the maximal set of
experimental data on pion, including the MSR and the decay
constant simultaneously, at the same values of the parameters of
the model \cite{Kru97}.

We use the following explicit form of the quark form factors:
$$
G^q_E(Q^2)=e_qf_q(Q^2),
$$
\[
G^q_M(Q^2)=(e_q+\kappa_q)f_q(Q^2),
\]
where $e_q$ are quarks charges and $\kappa_q$ --- anomalous magnetic
moments which enter our equations through the sum
 $s_q =\kappa_u+\kappa_{\bar d}$.
\begin{equation}
f_q(Q^2)=\frac{1}{1+\ln(1+\langle r_q^2\rangle Q^2/6)}\;,
\label{fq}
\end{equation}
where $\langle r^2_q\rangle$ is the quark MSR.

Let us discuss in brief the motivation for choosing the explicit
form (\ref{fq}). One of the features of our approach is the fact
that the form factor asymptotic behavior at
$Q^2\to\infty$, $M\to 0$ does not depend on the choice of the wave
function in Eq.~(\ref{ffpi}) and is defined by the relativistic kinematics
of two--quark system only \cite{KrT01}. In the point--like quark
approximation ($\kappa_q$=0, $\langle r^2_q\rangle$= 0)  the asymptotics
coincides with that described by quark counting laws \cite{MaM73,BrF73}
(see also the recent discussion in \cite{Rad09}): $F_\pi(Q^2) \sim
Q^{-2}$. The asymptotic behavior of the pion form factor was considered in
\cite{FaJ79,EfR80} (see also Ref.~\cite{Rad04}). The form (\ref{fq}) gives
logarithmic corrections to the power--law asymptotics , obtained in QCD.
So, in our approach the form (\ref{fq}) for the quark form factor gives
the same asymptotics as is predicted by QCD. Let us note that another
choice of the form of quark form factor, for example, the monopole form
\cite{Car94}, changes essentially the pion form factor asymptotics so that
it does not correspond to QCD asymptotics anymore.

To calculate the pion form factor we use wave functions of different forms:
harmonic oscillator wave functions (analogous to those used in the seminal
paper
\cite{ChC88pl} continued recently in \cite{CoP05}),
power-law-type wave functions with the explicit form motivated by
perturbative QCD calculations at high $Q^2$ \cite{Sch94,LeB80,BrL89}, wave
functions with linear confinement and Coulomb-like behavior at small
distances \cite{Tez91}. These functions are of the form:
$$
u(k)= N_{HO}\,\hbox{exp}\left(-{k^2}/{2b^2}\right)\;,
$$
\begin{equation}
 N_{HO} =
\sqrt{\frac{4}{\sqrt{\pi}\,b^3}}\;, \label{HO-wf}
\end{equation}

$$
u(k) = N_{PL}\,{(k^2/b^2 + 1)^{-n}}\;,\quad n = 3\;,
$$
\begin{equation}
N_{PL} =
16\sqrt{\frac{2}{7\,\pi\,b^3}}\;,
 \label{PL-wf}
\end{equation}

$$
u(r) = N_T \,\exp(-\alpha r^{3/2} - \beta r)\;,
$$
$$
\alpha = \frac{2}{3}\sqrt{2\,M_r\,a}\;,\;\beta = M_r\,b\;, \;N_T
= \frac{3\sqrt{2}\,\alpha}{\sqrt{N(\alpha\,,\beta})}\;,
$$
$$
\quad N(\alpha\,,\beta) =
9\,\alpha\,\sqrt[3]{2\alpha}\,\Gamma\left(\frac{5}{3}\right)\,{}_1F_1\left(\frac{5}{6},\,\frac{2}{3},\,t\right)
$$
$$
-2\,\sqrt[3]{4\,\alpha^2}\,\beta\,\Gamma\left(\frac{1}{3}\right){}_1F_1\left(\frac{7}{6},\,\frac{4}{3},\,t\right)
$$
\begin{equation}
+
6\,\beta\,^2\,{}_2F_2\left(1,\,\frac{3}{2},\,\frac{4}{3},\,\frac{5}{3},\,t\right)\;,\;
t =  -\,\frac{8\,\beta^3}{27\,\alpha^2}\;,
\label{Tez91-wf}
\end{equation}
where $a\;,b$ are the parameters of linear and Coulomb parts of potential,
respectively,
$M_r$ is the reduced mass of the two-particle system,
$b=(4/3)\,\alpha_s$,$\;\alpha_s= $0.59 on
the scale of the light mesons mass, ${}_pF_q$ are hypergeometric
functions, $\Gamma(x)$ is the Euler gamma function.

The parameters of the model are the same as in
Ref.~\cite{KrT01} where the motivation of the choice is described in
detail.
Let us note that for the constituent-quark mass
$M=0.22$ GeV  the values of parameters
(\ref{HO-wf}) -- (\ref{Tez91-wf})
were chosen in such a way as to ensure the pion MSR within experimental
uncertainties
$\langle r^2_\pi\rangle^{1/2}_{exp}$ = 0.657$\pm$0.012 fm
\cite{Ame84} as well as the best description of the decay constant
$f_{\pi\,exp}$ = 0.1317$\pm$ 0.0002 GeV \cite{PDG92}.
The sum of the quark anomalous magnetic moments is $s_q=$ 0.0268,
the quark MSR is $\langle r^2_q\rangle\;\simeq\,0.3/M^2$.
The values of other parameters are the following:
in the model (\ref{HO-wf}) $b=$0.3500 GeV (the decay constant is
$f_\pi=$ 127.4 MeV); in the model (\ref{PL-wf}) $n= 3\,,\,b=$0.6131
GeV ($f_\pi=$ 131.7 MeV); in the model (\ref{Tez91-wf})
$a=$0.1331 GeV$^2$ ($f_\pi=$ 131.7 MeV).
An interesting feature of our results is the fact that at the fixed
constituent quark mass, the dependence of the pion form factor on the
choice of the model (\ref{HO-wf}) -- (\ref{Tez91-wf}) is rather weak. The
curves calculated with different wave functions but one and the same quark
mass form groups \cite{KrT01}. From the theoretical point of view this
 weak dependence of our calculations on the model is the consequence of
the dispersion-relation base of the approach.

\section{RESULTS OF THE CALCULATIONS}

\begin{figure}
\begin{center}
\includegraphics[width=0.95\columnwidth]{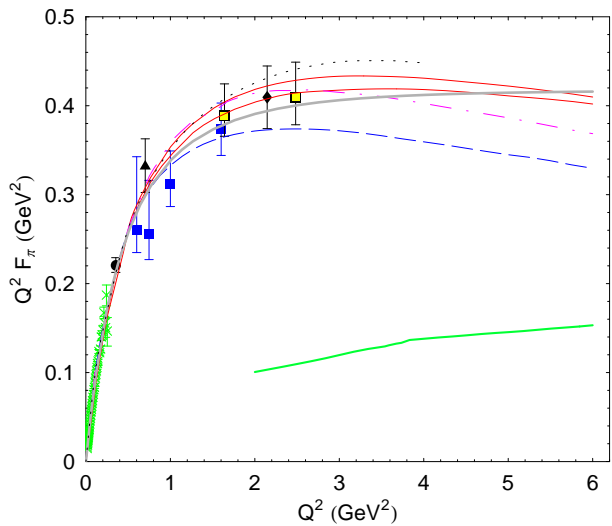}
\end{center}
\caption{
\label{fig1}
(color online).
Our predictions for the pion form factor given in 1998~\cite{KrT01}
(full red
lines; upper line: wave functions (\ref{HO-wf}), lower line: wave
functions (\ref{PL-wf}) with $n=3$ and (\ref{Tez91-wf})) compared with data
and with some other models. Green crosses represent data points of
Amendolia et al.~\cite{Ame84}. Other data points (all taken from
Ref.~\cite{HuB08-2}) are: reanalyzed points of Ackerman et al.
\cite{Ack78}
(full circle); reanalyzed points of Brauel et al. \cite{Bra79} (full
triangle); JLab results (full diamond, blue and yellow squares).
Other theoretical curves are: QCD approximation of
Maris and Tandy \cite{MaT00} (dotted);
perturbative QCD (leading and next-to-leading order) by Bakulev {\it et
al.} \cite{BaP04} (green dash-dot-dotted); Nesterenko
and Radyushkin \cite{NeR82} (magenta dash-dotted); dispersion
approach of Donoghue and Na \cite{DoN97} (blue dashed);
holographic approach of Grigoryan and Radyushkin \cite{GrR08} (thick
grey).}
\end{figure}

\begin{figure}
\begin{center}
\includegraphics[width=0.95\columnwidth]{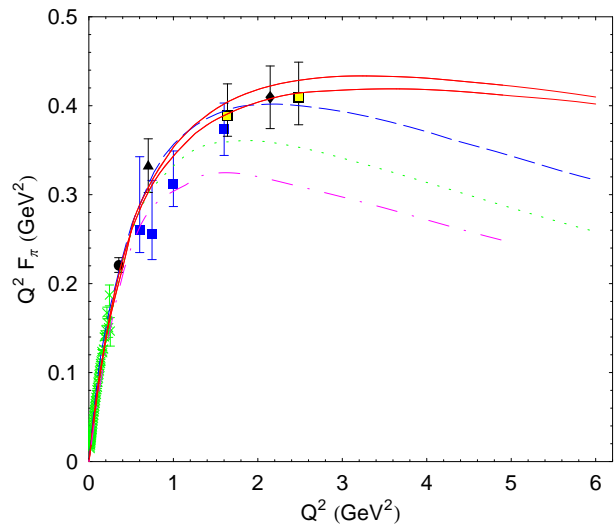}
\end{center}
\caption{
\label{fig2}
(color online).
Comparison of our predictions with other CQMs. Data and our curves are the
same as in Fig.~\ref{fig1}. Other theoretical curves are:
those by C.-W.~Hwang \cite{Hwa01} (blue dashed);
Cardarelli {\it et al.} \cite{Car94} and
(precisely coinciding with it) instant-form predictions \cite{HeJ04}
(magenta dash-dotted); Ref.~\cite{ChJ99} (green dotted). Predictions of an
upgraded version of a seminal paper \cite{ChC88pl} (Ref.~\cite{CoP05},
Fig.~5, $m_q = 0.22$) coincide precisely with our upper curve. }
\end{figure}

The results of the calculation of the charge pion form factor using the
wave functions (\ref{HO-wf}), (\ref{PL-wf}), (\ref{Tez91-wf}) and the
value of constituent-quark mass $M$ = 0.22\,GeV (this parameter has been
fixed as early as in 1998 \cite{KrT01}  from the data at $Q^2 \le 0.26$
(GeV)$^2$ \cite{Ame84}) are shown in Figures~\ref{fig1} and \ref{fig2}.
Let
us note that our relativistic CQM describes well the experimental data for
the pion form factor including the recent points \cite{HuB08-2}. The upper
of our curves corresponds to the model (\ref{HO-wf}), the lower - to the
models (\ref{PL-wf}) with $n= 3$ and (\ref{Tez91-wf}), which lie close to
one another.

Let us emphasize that the parameters used in our calculations were
obtained from the fitting to the experimental data up to
$Q^2\;\simeq$ 0.26 GeV$^2$ \cite{Ame84}. At that time the data for higher
$Q^2$ was not correlated in different experiments and had significant
uncertainties. The later data for pion form factor in JLab experiments up
to $Q^2$ =2.45 GeV$^2$ were obtained with rather good accuracy. All
 experimental points obtained in JLab up to now agree very well with our
prediction of 1998.

Let us discuss briefly the basic moments that provide good results for
the pion form factor in our approach. First, throughout the calculation the
condition of Lorentz-covariance and the conservation laws for the operator
of electromagnetic current were satisfied. Second, the condition of
accurate description of the pion MSR constrains the behavior of the
wave functions in momentum representation (\ref{HO-wf}), (\ref{PL-wf}) at
small relative momentum of quarks or of the wave function in coordinate
representation (\ref{Tez91-wf}) at large distances because of special
properties of the integral representation (\ref{ffpi}). Third, the
condition of the best description of the decay constant defines
constraints for the wave function at large relative momenta because the
contribution of small relative momenta to the decay constant is suppressed
as can be seen from the relativistic formula (see, e.g.
\cite{KrT01,Jau91}):
\[
f_\pi = \frac{M\,\sqrt{3}}{\pi}\,\int\,\frac{k^2\,dk}{(k^2 +
M^2)^{3/4}}\, u(k)\;.
\]

So, our way of fixing the model parameters constrains effectively the
behavior of wave functions both at small and at large relative momenta.
The structure of our relativistic integral representation (\ref{ffpi})
is so, that  the form--factor behavior in the region of small momentum
transfers is determined by the wave function at small relative momenta,
and the behavior of the form factor in the region of high momentum
transfer --- by the wave function at large relative momenta. The
constraints for the wave functions provide the limitations for the form
factor, and this is seen in the results of the calculation.

\section{ASYMPTOTIC BEHAVIOR}
It is worth to consider the form-factor
asymptotic behavior at
$Q^2\to\infty$ especially. In our paper \cite{KrT98tmph}  it was shown
that
in our approach, the pion form-factor asymptotics at $Q^2\to\infty$ and ¨
$M\to$ 0 does not depend on the choice of a wave function but is defined by
the relativistic kinematics only. We consider the fact that the asymptotics
obtained in our nonperturbative approach does coincide with that predicted
by QCD as a very significant one. Our approach occurs to be
consistent with the asymptotic freedom, and this feature surely
distinguishes it from other nonperturbative approaches.

Let us note that it is obvious that at very high momentum transfers the
quark mass decreases as it goes to zero at the infinity. Our approach
permits to take into account the dependence $M(Q^2)$ beginning from the
range where this becomes necessary to correspond to experimental data. It
is possible that this will take place at the values of $Q^2$ lower than
$6$ GeV$^2$.

The correct asymptotics is the consequence of the fact that the
relativism is an intrinsic property of our approach. To demonstrate how
it works let us consider the simple example of point--like quarks and
model wave functions (\ref{HO-wf}). In this case in the Eqs.
(\ref{ffpi}), (\ref{g_0}):
$$
G^u_E(Q^2)+G^{\bar{d}}_E(Q^2)=
$$
\[
= G^u_M(Q^2)+G^{\bar{d}}_M(Q^2) =
1\;.
\]

For the model (\ref{HO-wf}) it is easy to obtain the non-relativistic
integral representation of the form-factor as the corresponding limit of
the Eq.(\ref{ffpi}). Now the integration can be performed analytically
 and the following form for the nonrelativistic pion form factor can be
derived:
\[
F_\pi(Q^2) = \hbox{exp}\left(-\,\frac{Q^2}{16\,b^2}\right)\;.
\]

One can see that in the non-relativistic case, the form factor does not
depend on the mass of constituents and its asymptotics can not agree with
that of QCD. The correct asymptotics is provided by relativistic effects.

In the relativistic case the results for the integrals can not be obtained
analytically. To derive the asymptotic behavior in question it is possible
to use the asymptotic series for double integrals obtained in
Ref.~\cite{KrT08}. The first two terms give:
\begin{equation}
F_\pi(Q^2)\sim\frac{2^{5/2}M}{Q}\mathrm{e}^{-\frac{QM}{4b^2}}\left(1+\frac{7b^2}{2MQ}\right)\;.
\label{2bes}
\end{equation}

Let us take in (\ref{2bes}) the limit at $M/b\to 0$.
This means that the parameters of the model are such that
$M/b\ll 1$. The physical meaning of this limit is that the increase of the
momentum transfer is followed by the ``undressing'' of the constituent
quarks and its transformation into current quark of pQCD.
In this limit we obtain from
Eq.~(\ref{2bes}) up to logarithmic prefactors the power-like behavior
 coinciding with that of pQCD
~\cite{KrT01}:
\[
F_\pi(Q^2)\sim\frac{14\sqrt{2}b^2}{Q^2}\;,
\]

\section{SUMMARY}

To conclude, we make some predictions about the results of the future JLab
experiments on the pion form factor based on the method proposed in our
papers earlier. The method is a variant of composite quark model in the
instant-form of Poincar\'e invariant quantum mechanics. It is shown that
our approach has certain advantages as compared with other CQM
calculations. From a theoretical point of view, these advantages are the
consequence of the fact that our approach has dispersion-relation
motivated foundations. This provides, in particular, weak model dependence
of the results of calculations. The approach has demonstrated
earlier its predictive power in describing all the data
on the pion form factor obtained later in JLab experiments. Our
calculations also give the accurate values of the pion MSR and of the decay
constant $f_\pi$, and the correct asymptotic behavior at $Q^2\to\infty$.

We hope that our model will provide a good description of the future JLab
experiments on the measurement of the pion form factor in the range of
momentum transfers up to $Q^2 \approx 6$ (GeV)$^2$.
If it will occur that
beginning from some values one needs to adjust the calculations by
introducing the quark mass dependence on $Q^2$ to agree the future data,
then we propose to identify the effect with the appearance of
perturbative effects.

This work was supported in part by
the Russian Foundation for Basic Research (Grant No. 07-02-00962). The
work of A.K.\ was supported in part by the Program "Scientific and
scientific-pedagogical specialists of innovative Russia" (Grant
No.~1338).




\begin{thebibliography}{49}

\bibitem{Blo08-1} H.P.~Blok  {\it et al.} (The Jefferson Lab $F_\pi$
Collaboration), Phys. Rev. C \textbf{78}, 045202 (2008).
\bibitem{HuB08-2} G.M.~Huber {\it et al.} (The Jefferson Lab $F_\pi$
Collaboration), Phys. Rev. C \textbf{78}, 045203 (2008).
\bibitem{Proposal06} G.M.~Huber, D.~Gaskell,
                     JLab Proposal, E-12-06-101,
                     ``Measurement of the charged pion form factor
                     to high $Q^2$'', July 7, 2006.
\bibitem{BlH02} H.P.~Blok, G.M.~Huber, and D.J.~Mack,
                 ``Measurement of th¥ Charged Pion Form Factor to High
                 $Q^2$ '',  Contribution to Exclusive Reaction Workshop,
                 Jefferson Lab, May 2002,\\
nucl-ex/0208011
\bibitem{EbF06} D. Ebert, R.N. Faustov, and V.O. Galkin, Eur.Phys.J.C
\textbf{47}, 745 (2006).
\bibitem{KrT01} A.F.~Krutov and V.E.~Troitsky,
                 Eur.Phys.J. C \textbf{20}, 71 (2001), hep-ph/9811318.
\bibitem{Vol01}  J.~Volmer {\it et al.},
                 Phys. Rev. Lett. {\bf 86}, 1713 (2001).
\bibitem{Tad07}  V.~Tadevosyan {\it  et al.},
                 Phys. Rev. C {\bf 75}, 055205 (2007).
\bibitem{Hor06} T.~Horn {\it et al.},
                Phys. Rev. Lett.{\bf  97}, 192001 (2006).
\bibitem{KrT02}  A.F.~Krutov and V.E.~Troitsky,
                 Phys.Rev. C \textbf{65}, 045501 (2002).
\bibitem{KrT01long} A.F.~Krutov and V.E.~Troitsky,
                hep-ph/0101327.
\bibitem{KrT09} A.F.~Krutov and V.E.~Troitsky,
                Fiz. Elem. Chastits At. Yad. \textbf{40}, 269 (2009)
                [[Engl. Transl. Physics of Particles and Nuclei,
                 \textbf{40}, 136 (2009)].
\bibitem{Ame84} S.R.~Amendolia {\it et al.},
                Phys. Lett. {\bf B146},116 (1984).\\
                 S.R.~Amendolia {\it et al.},
                 Nucl. Phys. {\bf B277}, 168 (1986).
\bibitem{DeD08} B.~Desplanques and Y.B.~Dong,
                Eur. Phys. J. {\bf A37}, 433 (2008).
\bibitem{ShT69} Yu. M.~Shirokov and V. E.~Troitsky,
                Nucl.Phys.B \textbf{10}, 118 (1969).
\bibitem{KoT72}  V. P.~Kozhevnikov, V. E.~Troitsky, S. V.~Trubnikov, and
                 Yu. M.~Shirokov, Teor. Mat. Fiz. \textbf{10}, 47 (1972)
                 [Engl. Transl.
                 Theor. Math. Phys.  \textbf{10}, 30 (1972)].
\bibitem{TrT72}  V.E. Troitsky, S. V. Trubnikov, and Yu. M.
                 Shirokov, Teor. Mat. Fiz. \textbf{10}, 209 (1972) [Engl.
                 Transl. Theor. Math. Phys.  \textbf{10}, 136 (1972)]
                  V.E. Troitsky, S. V. Trubnikov, and Yu. M. Shirokov,
                  Teor. Mat. Fiz. \textbf{10}, 349 (1972) [Engl. Transl.
                  Theor. Math. Phys.  \textbf{10}, 234 (1972)]
\bibitem{KiT75}  A. I.~Kirillov , V. E.~Troitsky , S. V.~Trubnikov , and
                 Yu. M.~ Shirokov, Fiz. Elem. Chastits At. Yad. \textbf{6}
                 ,3 (1975) [[Engl. Transl.  Sov. J. Part. Nucl. \textbf{6},
                 3 (1975)].
\bibitem{Tro93}  V. E.~Troitsky, in Quantum Inversion Theory and Applications,
                 Proceedings, Germany, 1993, edited by H.V.von
                 Geramb, Lecture Notes in Physics 427, 50 (Springer
                 Verlag, 1994).
\bibitem{AnK92}  V. V.~Anisovich, M. N.~Kobrinsky, D. I.~Melikhov, and
                 A .V.~Sarantsev, Nucl. Phys. B544 747 (1992).
\bibitem{Mel02}  D. I. Melikhov,  Eur. Phys. J. direct C4 (2002) 2
                 [hep-ph/0110087].
\bibitem{KrT99}
A.F.~Krutov and V.E.~Troitsky, JHEP {\bf 10}, 028 (1999).
\bibitem{Kru97}
A.F.~Krutov, Yad. Fiz. {\bf 60}, 1442 (1997) [Phys. At. Nuclei
{\bf 60}, 1305 (1997)].
\bibitem{MaM73} V.A.~Matveev, R.M.~Muradyan, and A.N.~Tavkhelidze,
                Lett. Nuovo Cim. \textbf{7} 719 (1973).
\bibitem{BrF73} S.~Brodsky and G.~Farrar,
                Phys.Rev.Lett. \textbf{31} 1153 (1973).
\bibitem{Rad09} A.~Radyushkin,
                ''Quark Counting Rules: Old and New Approaches'',
                arXiv:0907.4585.
\bibitem{FaJ79}  G. R. Farrar and D. R. Jackson,
                 Phys. Rev. Lett. \textbf{43}, 246 (1979).
\bibitem{EfR80}  A. V. Efremov and A. V. Radyushkin,
                 Teor. Mat. Fiz. \textbf{42}, 147 (1980) [Engl. Transl.
                 Theor. Math. Phys.  \textbf{42}, 97 (1980)]
                 Report JINRE2-11983, Oct. 1978, 32pp.
\bibitem{Rad04}  A. V. Radyushkin, hep-ph/0410276  JLAB-THY-04-35 October
                 20, 2004, JINR P2-10717 June 14,1977
\bibitem{Car94} F.~Cardarelli {\it et al.},
                Phys. Lett. {\bf B332}, 1 (1994);
                F.~Cardarelli {\it et al.},
                Phys. Rev. D {\bf 53}, 6682 (1996).
\bibitem{ChC88pl} P.L.~Chung. F.~Coester, and W.N.~Polyzou,
                Phys. Lett.{\bf B205} 545 (1988).
\bibitem{CoP05} F.~Coester and W.N.~Polyzou,
                Phys. Rev. C \textbf{71}, 028202 (2005).
\bibitem{Sch94} F.~Schlumpf,
                Phys. Rev. D \textbf{50}, 6895 (1994).
\bibitem{LeB80}
                G.P.~Lepage, S.J.~Brodsky,
                Phys.Rev.D \textbf{22}, 2157 (1980).
\bibitem{BrL89}
               S.J.~Brodsky, G.P.~Lepage,
               Perturbative Quantum Chromodynamics. -
               Singapore: World Scientific Publishing, 1989. - 93 p.
\bibitem{Tez91} H.~Tezuka,
                J. Phys. A Math. Gen. \textbf{24}, 5267 (1991).
\bibitem{PDG92}
                Particle Data Group. Part II,
                Phys.Rev.D \textbf{45} (1992).
\bibitem{Ack78} H.~Ackermann {\it et al.},
                Nucl. Phys. {\bf B137},294 (1978).
\bibitem{Bra79} P.~Brauel {\it et al.},
                Z. Phys. {\bf C3}, 101 (1979).
\bibitem{MaT00} P.~Maris and P.C.~Tandy,
                Phys. Rev. C {\bf 62}, 055204 (2000).
\bibitem{BaP04} A.P. Bakulev, K. Passek-Kumericki, W. Schroers, and
                N. G. Stefanis, Phys. Rev. D {\bf 70}, 033014 (2004), and
                erratum Phys. Rev. D {\bf 70}, 079906 (2004)
\bibitem{NeR82} V.A.~Nesterenko and A.V.~Radyushkin,
                Phys. Lett. {\bf B115}, 410 (1982).
\bibitem{DoN97} J.F.~Donoghue and E.S.~Na,
                Phys. Rev. D {\bf 56}, 7073 (1997).
\bibitem{GrR08} H. V.~Grigoryan and A. V.~Radyushkin,
                Phys. Rev. D {\bf 78}, 115008 (2008).
\bibitem{Hwa01} C.-W.~Hwang,
                Phys. Rev. D {\bf 64}, 034011 (2001).
\bibitem{HeJ04} Jun~He, B.~Juli\'a-D\'{\i}az, and Yu-bing~Dong,
                Phys.Lett. {\bf B602}, 212 (2004).
\bibitem{ChJ99} H.-M.~Choi and C.-R.~Ji,
                Phys. Rev. D {\bf 59} 074015 (1999).
\bibitem{Jau91} W.~Jaus,
                Phys.Rev.D \textbf{44}, 2851 (1991).
\bibitem{KrT98tmph} A. F.~Krutov and V. E.~Troitsky,
                Teor. Mat. Fiz. \textbf{116}, 215 (1998) [Engl. Transl.
                Theor. Math. Phys.  \textbf{116}, 907 (1998)].
\bibitem{KrT08} A.F.~Krutov, V.E.~Troitsky and N.A.~Tsirova,
                J. Phys. A: Math. Theor. \textbf{41}, 255401 (2008).

\end{thebibliography}
\end{document}